\documentclass{PoS}

\usepackage{amsmath}
\usepackage{cite}

\newcommand{\mys}[1]{\section{#1} \setcounter{equation}{0}}

\newlength{\dummysp}
\newcommand{\tr}{\mathop{{\hbox{Tr} \, }}\nolimits}

\newcommand{\beq}{\begin{eqnarray}}
\newcommand{\eeq}{\end{eqnarray}}
\newcommand{\nnn}{ \nonumber \\ }
\newcommand{\p}{{\partial}}

\newcommand{\e}{{\epsilon}}
\newcommand{\s}{{\sigma}}

\newcommand{\gappeq}{\mathrel{\rlap {\raise.5ex\hbox{$>$}}
{\lower.5ex\hbox{$\sim$}}}}
\newcommand{\lappeq}{\mathrel{\rlap{\raise.5ex\hbox{$<$}}
{\lower.5ex\hbox{$\sim$}}}}
\newcommand{\myref}[1]{(\ref{#1})}

\newcommand{\bfe}[1]{\vspace{4pt} {\bf #1 \hspace{2pt}}}

\newcommand{\ben}{\begin{enumerate}}
\newcommand{\een}{\end{enumerate}}

\newcommand{\phib}{{\bar \phi}}

\newcommand{\hc}{{\rm h.c.}}

\newcommand{\bit}{\begin{itemize}}
\newcommand{\eit}{\end{itemize}}

\newcommand{\susy}{supersymmetry}
\newcommand{\susyc}{supersymmetric}
\newcommand{\obf}{{\bf 1}}

\newcommand{\Ncal}{{\cal N}}

\newcommand{\Phib}{{\bar \Phi}}

\def\[{\left [}
\def\]{\right ]}
\def\({\left (}
\def\){\right )}

\title{Advances and applications of lattice supersymmetry}
\ShortTitle{Lattice supersymmetry}
\author{Joel Giedt \\ 
Fine Theor.~Phys.~Inst., Univ.~of Minnesota,
116 Church St.~S.E., Minneapolis, MN 55455 USA \\
E-mail:  \email{giedt@physics.umn.edu}}

\abstract{I review motivations for the study of supersymmetric
field theories by lattice techniques.  In particular, some of the
more interesting potential applications are described.  These
are models of quantum gravity, that rely on the AdS/CFT correspondence.
The dual gauge theory is an ideal place for lattice studies to
make a contribution.  I also survey some of the recent
advances in lattice formulations of supersymmetric theories.}

\FullConference{{\rm Plenary talk given at:} \\
XXIV International Symposium on Lattice Field Theory\\
July 23-28 2006 \hfill {\rm FTPI-MINN-06/35} \\
Tucson Arizona, US \hfill {\rm UMN-TH-2522/06}}

\begin{document}

\mys{Introduction}
Several motivations exist for efforts to 
formulate supersymmetric field theories on a lattice.
As is well known, it is difficult to formulate these theories in
a way that avoids fine-tuning of counterterms.
Nevertheless, there have been many promising
formulations developed of late.\footnote{For
recent reviews, see \cite{Giedt:2006pd,Catterall:2005eh,Feo:2004kx,
Feo:2002yi,Montvay:2001aj}.}
But, given a ``good'' formulation,
to what use should or would we put it?
To answer this question, let us look to the
motivations for putting \susyc\ theories on
the lattice.

\mys{To provide a definition of the field theory}
\subsection{Some general remarks}
Here, one wants to say in a precise way what is meant by
the rather formal expression
\beq
Z[S] = \int \exp ( -S ),
\eeq
the Euclidean functional integral, or partition
function for the system.
At a formal level we know that it is described
by imposing an ultraviolet (UV) cutoff length scale $a$,
and defining the integral for each $a$:
\beq
Z[S;a] = \int_{{\cal D}(a)} [d\mu(\phi;a)] \exp(-S[\phi;a]).
\eeq
Here, $[d\mu]$ is the integration measure and
${\cal D}$ is the domain, both of which might
contain explicit $a$-dependence.
This leads to an $a \to 0$ limiting sequence
that defines the physical theory.
In the lattice formulation, $a$ is of course
the lattice spacing, and $a \to 0$ is the
continuum limit.

Due to universality, there is no unique definition
of $Z[S;a]$.  That is, two definitions that differ
by irrelevant operators can be made to agree in
their continuum limit.
But among various proposals for $Z[S;a]$,
one wants a classification of the
extent to which local counterterms (CTs)
need to be adjusted to obtain the correct
continuum limit.  The best definitions
of $Z[S;a]$ possess the continuum symmetries
(e.g., the
exact gauge symmetry of Wilson's action),
or have lattice symmetries that guarantee
the continuum symmetries are recovered
as $a \to 0$ (e.g., the lattice chiral symmetry
of Ginsparg-Wilson fermions).  In either
case, the continuum symmetry results
without any adjustment of CTs.
In some cases, the lattice theory
explicitly breaks continuum symmetries,
and the symmetries are only obtained
through a fine-tuning of CTs.
It may even be that
no local CTs exist that will do the job,
in which case we have an anomaly.  We should
study the question of CTs both perturbatively
and nonperturbatively.  Our favorite answer
of course would be that no symmetry-restoring 
CTs are required.\footnote{Of course,
CTs associated with renormalization are
required unless the theory is finite.
But these are CTs allowed by the continuum symmetry
and will not concern us here.}

\subsection{Examples without CTs}
I will now enumerate several examples where
no CTs are required.  The point here is
to familiarize the reader with the
many successes, or claimed successes, that have been achieved
in this regard.  Some of these are quite
recent.

\subsubsection{Q-exact \susyc\ quantum mechanics}
In this case, one has a lattice version of a supercharge $Q$
that is nilpotent; i.e. $Q^2=0$.
The lattice action is describable
as $Q$ acting on something, ``$X$'':
$S=QX,$ and is trivially invariant:  $QS=Q^2X=0$.

There is convincing Monte Carlo evidence 
that the lattice theory in this case
has the correct continuum limit
without fine-tuning of CTs.  This
evidence consists of Bose-Fermi spectrum degeneracy and
\susy\ Ward identities \cite{Catterall:2000rv}.
For example, in Table~\ref{cgre}
one sees that the mass gap for fermionic
states is degenerate with that for
bosonic states, even at finite lattice spacing.

An all orders perturbative proof of the
absence of CTs in this theory was given
in \cite{Giedt:2004qs}.  Moreover, a
nonperturbative proof of the absence of CTs
was also given in \cite{Giedt:2004qs}, using transfer 
matrix methods.
In either calculation, it is
apparent that the Q-exact nature
of the action leads to cancellations 
that avoid the appearance of CTs.
These cancellations are associated with
the exact \susy\ Ward identities that
follows from $QS=0$.

\begin{table}
\begin{center}
\begin{tabular}{ccc}
\hline
$N$  & $m_B$      &$m_F$      \\ \hline
16 & $10.60(3)$ & $10.64(5)$ \\
32 & $12.80(2)$ & $12.91(4)$ \\
64 & $14.47(2)$ & $14.52(2)$ \\
128 & $15.63(4)$ & $15.63(4)$ \\
256 & $16.19(3)$ & $16.28(4)$ \\ \hline
\end{tabular}
\end{center}
\caption{
Spectrum degeneracy in Q-exact \susyc\ quantum
mechanics.  These results are from \cite{Catterall:2000rv}.
Note that with $L=Na=$ fixed, increasing
$N$ corresponds to the continuum limit.
The $N \to \infty$ extrapolation matches
the (numerical) solution of the continuum
Schr\"odinger equation.
\label{cgre}
}
\end{table}

\subsubsection{Q-exact 2d $\Ncal=2$ Wess-Zumino (a.k.a.~$\Ncal=2$ Landau-Ginsburg)
model}
\begin{figure}
\begin{center}
\includegraphics[width=4in,height=3in,bb=65 400 500 750,clip]{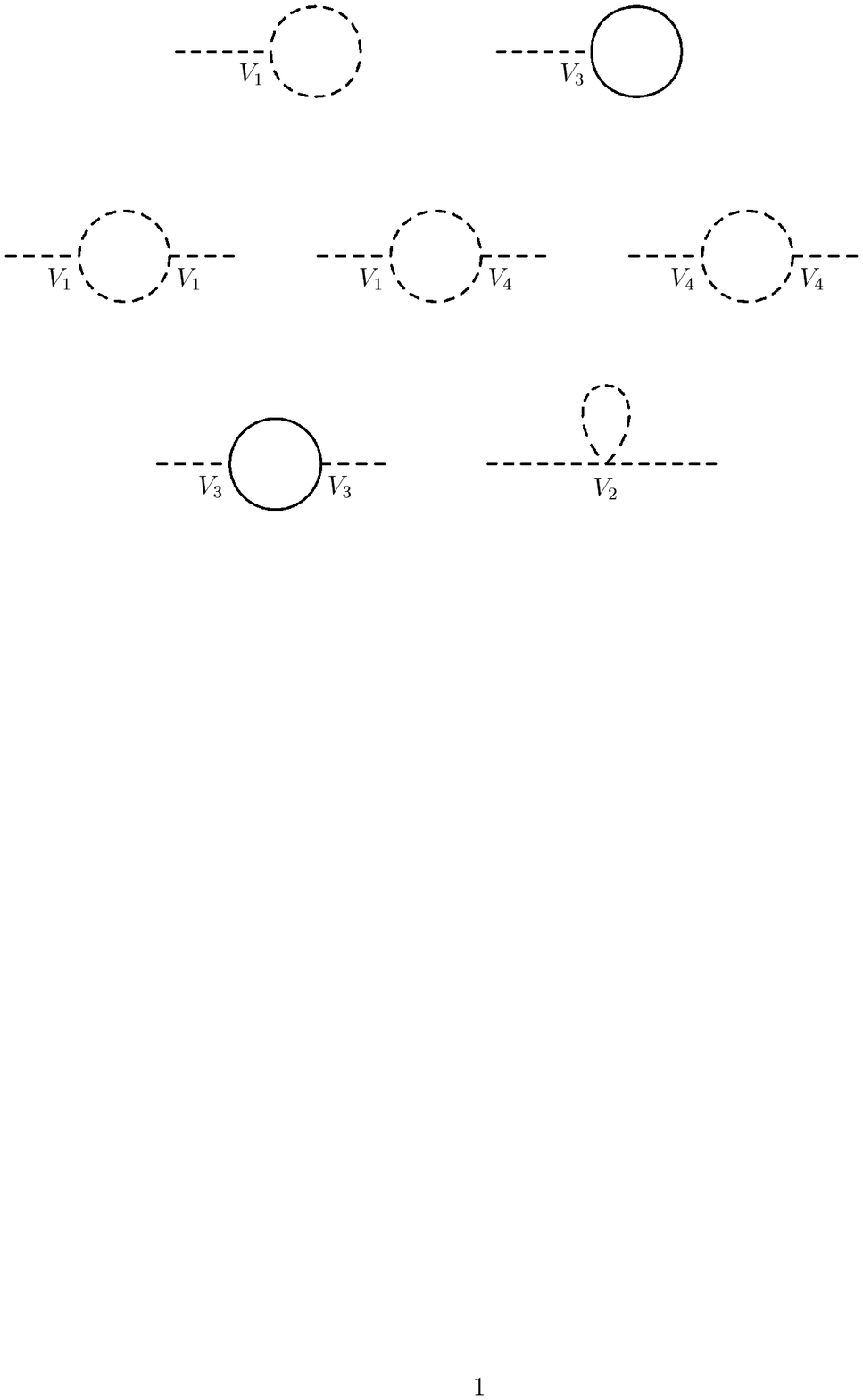}
\end{center}
\caption{The set of all 1PI diagrams with UV degree of divergence
$D_{UV} \geq 0$ in the Q-exact lattice 2d $\Ncal=2$ Wess-Zumino model. 
The notation for the vertices, which are modified
on the lattice, is explained in \cite{Giedt:2004qs}. 
\label{2wzf}}
\end{figure}

Here, the model is the 4d $\to$ 2d dimensional
reduction of the Wess-Zumino model with a cubic
superpotential \cite{Wess:1973kz}.
The Q-exact lattice formulation of this model was
given long ago \cite{Elitzur:1983nj,Sakai:1983dg}.
An all orders perturbative proof of the
absence of CTs was given in \cite{Giedt:2004qs}.
That proof is similar to the one for the
$Q,Q^\dagger$-preserving spatial lattice \cite{Elitzur:1983nj}.
In both cases, Q-exactness leads to \susy\
Ward identities, which cause the cancellations
that avoid CTs.  Fig.~\ref{2wzf} shows all the
diagrams that must be studied (the lattice theory
is super-renormalizable).  The parts that
would potentially differ from the continuum
perturbation series, according to Reisz's 
theorem \cite{Reisz:1987da}, all cancel.

On the nonperturbative front, there is
various numerical evidence (spectrum degeneracy, Ward
identities, R-symmetry) that the
\susyc\ limit is obtained without
CTs \cite{Beccaria:1998vi,Catterall:2001fr,Catterall:2001wx,
Giedt:2005ae}.

\subsubsection{$\Ncal=1$ 4d super-Yang-Mills with Ginsparg-Wilson fermions}
From the work of \cite{Curci:1986sm}, we know that
$\Ncal=1$ 4d super-Yang-Mills (SYM) with Ginsparg-Wilson fermions
requires no \susy-restoring CTs.  This is because the chiral limit and
the \susyc\ limit in this model are equivalent:
the only gauge invariant relevant operator
that could break \susy\ is a gaugino mass.
The application of overlap fermions 
\cite{Narayanan:1992wx,Narayanan:1993sk,Neuberger:1997fp}
to this model was sketched in \cite{Maru:1997kh,Neuberger:1997bg}.
But basic simulation studies,
such as glueball spectra, have yet to be attempted using
overlap fermions.
The application of domain wall fermions \cite{Kaplan:1992bt}
to this model was proposed \cite{Neuberger:1997bg,Nishimura:1997vg},
sketched \cite{Kaplan:1999jn} and briefly studied \cite{Fleming:2000fa}.
However, it deserves another push:  much more could
be accomplished with the computational resources that are
now available.\footnote{I thank Pavlos Vranas
for extensive conversations on what could
be achieved presently.} 
Infrared (IR) effective theories for this
model have been proposed by continuum
theorists, based on various arguments \cite{Veneziano:1982ah,
Farrar:1997fn,Farrar:1998rm,Gabadadze:1998bi,Cerdeno:2003us,Merlatti:2004df,
Auzzi:2005fi}.  In particular, the predicted IR spectrum of states
in the presence of a small gluino mass has a pattern of splitting
\cite{Farrar:1998rm} that would be
very interesting to test in a lattice
formulation that is sensitive to topology,
such as gap domain wall fermions \cite{Vranas:2006zk}.

\subsubsection{Deconstruction models}
Deconstruction models (a.k.a.~quiver lattice, orbifold matrix model)
offer another route that avoids some or all CTs.\footnote{These
models are inspired by earlier, phenomenological works on
dimensional deconstruction \cite{Arkani-Hamed:2001ca,Hill:2000mu}.}
These have been written down for
various extended SYM and super-QCD models
in 2d, 3d and 4d \cite{Kaplan:2002wv,Cohen:2003xe,Cohen:2003qw,
Kaplan:2005ta,Endres:2006ic,Giedt:2006dd}.
It is easily shown, on symmetry and dimensional
grounds, that the 2d SYM examples require no \susy-restoring
CTs in perturbation theory.  In the case of 2d super-QCD,
renormalization has yet to be explored.
For the $d>2$ examples, it is less clear what really
happens; explicit calculations will be required.
However, if the Euclideanized d-dimensional 
Poincar\'e group is recovered, an interesting
\susyc\ theory is obtained either way, since
there is at least some supersymmetry present,
even if not the full extended supersymmetry of the
target theory.\footnote{Thanks are due David Kaplan
for emphasizing this point during conversations.}
A review of the deconstruction models was recently given in
\cite{Giedt:2006pd}.  Perturbative
aspects of one of the models was
studied in \cite{Onogi:2005cz}.
In \cite{Giedt:2003ve,Giedt:2003vy,Giedt:2004tn},
the phase of the fermion determinant
was found to be problematic in two of the models.

\subsubsection{D'Adda et al.~SYM}
Extended SYM in 2d and 4d has been formulated
in a very interesting way by D'Adda 
et al.~\cite{D'Adda:2004jb,D'Adda:2005zk}.
It is based on a
modified Leibnitz rule for the \susy\ variations $s_A$
of products of fields:
\beq
&& s_A[\Phi(x_1) P(x_2,\ldots)] 
= s_A[\Phi(x_1)] P(x_2,\ldots) \nnn
&& \quad + ~ (-)^{F(\Phi)} \Phi(x_1+a_A) s_A[P_2(x_2,\ldots)].
\eeq
Note that the site index $x_1$ of $\Phi$ has
shifted as $s_A$ is pushed past it.  It
is in this sense that the formulation is
``slightly noncommutative.''

For the 2d (2,2)-\susyc\ case,
the action can be written in four equivalent forms:
\beq
S &=& \sum_x \tr s \tilde s s_1 s_2 \Psi_{x,x}
= - \sum_x \tr \tilde s s s_1 s_2 \Psi_{x,x} \nnn
&=& \sum_x \tr s_1 s_2 s \tilde s \Psi_{x,x} 
= - \sum_x \tr s_2 s_1 s \tilde s \Psi_{x,x},
\label{dasi}
\eeq
where $\Psi_{x,x}$ is a composite operator whose
exact form we will not need here.
Due to the nilpotentcy of the \susy\ operators $s_A$,
\beq
s^2 = {\tilde s}^2 = s_1^2 = s_2^2 = 0,
\label{nilly}
\eeq
the action is invariant under the modified (2,2) \susy.
The renormalization of the model needs more study.  
In particular, does the slight noncommutativity matter?

Bruckmann and de Kok have studied the noncommutativity approach
in supersymmetric quantum mechanics \cite{Bruckmann:2006ub}.  They find
that the existence of modified supersymmetric invariance depends on
the order in which the lattice fields are written.  Since
the order of the lattice fields defines the symmetry
transformation, this is just the statement
that the invariance (or lack of it) depends on how one
decides to transform the fields.
I do not necessarily regard this as an inconsistency of the
noncommutative approach, but rather as an illustration
that in some theories there may be an ordering ambiguity
that needs to be resolved in order to have the modified
supersymmetric invariance.  Indeed, Bruckmann and de Kok seem to find that
such an ordering does exist in the supersymmetric
quantum mechanics model.

By contrast, the ordering in the D'Adda et al.~approach is completely
determined by the lattice gauge invariance.  When the
action is written in the form \myref{dasi}, there is no
ambiguity resulting from the ordering of terms; $\Psi_{x,x}$,
appearing in \myref{dasi}, is unique.
The invariance of the component field action follows
identically from the properties \myref{dasi} and \myref{nilly}.
All of this is easily checked by straightforward,
if somewhat tedious, algebraic manipulations.

On the other hand, these are formal manipulations;
their concrete realization in terms of component fields
is an area of active study.  In \cite{Bruckmann:2006kb},
Bruckmann et al.~argue that, due to the link character
of most of the supersymmetry variations, their
properties on gauge invariant loops is not well-defined.
However, as a final remark they acknowledge that
the introduction of link-valued Grassmann parameters,
as suggested by D'Adda, could provide a way out.
Suffice to say, the D'Adda et al.~proposal deserves
further study.

\subsubsection{Other no CT examples/claims}
Other recent examples where there
is evidence presented that fine-tuning
of CTs is avoided include:
\bit
\item
Q-exact compact (2,2) SYM in 2d as formulated by Sugino in
\cite{Sugino:2003yb,Sugino:2004qd,Sugino:2004uv,Sugino:2006uf}.
There is no fine-tuning in perturbation theory.
\item
Twisted (Q-exact) geometrical (2,2) SYM in 2d, due
to Catterall \cite{Catterall:2004np}.
Monte Carlo data seems to indicate no need for CTs
\cite{Catterall:2006jw}.  Of practical
importance, the phase quenched ensemble
in this model is a good approximation to
the full ensemble, 
due to a sharply peaked phase distribution
as one approaches the continuum limit.
\item
The 4d Wess-Zumino model with Ginsparg-Wilson fermions,
recently proposed and studied by
Bonini and Feo \cite{Bonini:2004pm,
Bonini:2004wv,Bonini:2005qx}.
Here, there is a nonlinear, perturbative definition of 
the supercharges $Q_A$.  
\item
The twisted (Q-exact) nonlinear 2d $\s$ model with
(2,2) \susy\ \cite{Catterall:2003uf}.  A superfield
formulation was subsequently given in \cite{Giedt:2004qs},
where renormalization was discussed
on symmetry and dimensional grounds.
Whereas that study showed that
radiative generation of \susy\ violation 
in the continuum limit is not forbidden
by lattice symmetries and dimensional arguments, recent
Monte Carlo data seems to indicate no need for 
\susy-restoring CTs \cite{Ghadab:2004rt,Catterall:2006sj}.
\eit

In summary, much progress has been made
recently in formulating \susyc\ theories
on the lattice.  In some cases, the theory
has been nonperturbatively defined.  Furthermore,
numerical results have been obtained in
a few cases, holding out the prospect that
we will obtain interesting information by means other
than the traditional, continuum methods.

\section{Nonholomorphic woes}
Now that we have seen that there are
many promising formulations that define
\susyc\ theories on the lattice, let us
return to the motivations.
Continuum \susy\ tricks usually partly fail
to determine the IR effective theory.

Schematically, the IR effective theory
is described by an action of the form:
\beq
\int d^4\theta ~ K(\Phi,\Phib) + \[ \int d^2\theta ~
W(\Phi) + \hc \].
\eeq
Here, integration over Grassmann superspace
coordinates $\theta,\bar\theta$ has been
used, and $\Phi,\Phib$ are chiral superfields.  The superpotential
$W$ is not renormalized in perturbation theory.
Also, $W$ is sometimes completely determined, once
symmetries accounted for.
Generically none of these nice features hold for 
the K\"ahler potential $K$.

This lack of control over nonholomorphic data, encoded
in $K$, is
distressing.  This is because the
so-called ``supersymmetry-breaking soft-terms''
largely determine superpartner spectra
and couplings for the MSSM, and these
soft-terms depend on the K\"ahler potential $K$.
To the extent that nonperturbative physics
is involved in breaking supersymmetry,
the nonperturbative renormalization of
the K\"ahler potential is important to
the prediction of soft-terms.

Lattice Monte Carlo simulations would, as a first step,
give us a handle on vevs $\phi_0$ of scalars, 
and the spectrum of light states.
This constrains $\p V/ \p \phi$, $\p^2 V / \p \phi \p \phi$
and $\p^2 V / \p \phi \p \phib$
evaluated at $\phi_0$.  
Both the K\"ahler potential $K$ and superpotential $W$
play a role in the scalar potential,
$V=K^{k \bar \ell} W_k \bar W_{\bar \ell}$,
where $K^{k \bar \ell}$ is the inverse of the K\"ahler metric 
$K_{k \bar \ell} = \p^2 K/ \p \phi^k \p \bar\phi^{\bar \ell}$,
and $W_k = \p W/\p \phi_k, \;
\bar W_{\bar \ell} = (W_\ell)^*$.

A scheme for extracting the effective $K$ would be:
(1) Hypothesize an effective K\"ahler potential $K$,
containing some minimal set of phenomenological
constants.
(2) Use the known effective superpotential $W$.
(3) Match microscopic lattice and IR effective lattice data,
to fit ``phenomenological'' constants in $K$.
This outline illustrates how lattice simulations 
have the potential to teach us something
about nonperturbative renormalization of nonholomorphic quantities.

In work that is in progress with Simon Catterall,
we are attempting to determine the (2,2) 2d 
twisted nonlinear $\s$ model that is the
IR effective theory for (2,2) 2d SU(2) SYM.
For example, we are comparing constrained effective
potentials in the two theories, for
various scalars operators.

\section{Supersymmetry breaking}
A third motivation is to
improve our understanding of
dynamical supersymmetry breaking.
Strong \susy\ dynamics is often invoked in 
models of soft \susy\ breaking for the
Minimal Supersymmetry Standard Model (MSSM).
Any improvement of our understanding
of the strong dynamics of \susyc\ gauge
theories would be helpful.

As an example, consider 3d $\Ncal=2$ SYM.
This is a model that is perhaps feasible for
thorough lattice study.  
The theory has a
simple content (one gluon, one adjoint Majorana fermion, 
one adjoint real scalar), so it will be
efficient to simulate.
It is
a simple theory where the ground state \susy\ 
is not yet fully understood.
It is obtained from the 3d reduction of 
4d $\Ncal=1$ SYM.

This model was studied from the continuum
in \cite{Affleck:1982as}.
It was found that there is an
instanton-generated potential for the modulus field $\phi$.
However, there
is uncertainty regarding the potential
for small $\phi$, because the theory is strongly
coupled in that regime.
The only known \susyc\ vacuum is at $\phi \to \infty$.
We would like to know:  Is there a \susyc\ vacuum near origin
in field space?

It seems promising to 
use a parity-preserving overlap-Dirac formulation, similar
to 3d $\Ncal=1$ model of \cite{Maru:1997kh}.
If one does this, do we avoid fine-tuning,
as in the 3d $\Ncal=1$ case?  It
is a straightforward, open question to be
explored.  


\mys{Quantum gravity and the AdS/CFT correspondence}
There is an evolving understanding of the relationship
between SYM and string/M-theory.
A nonperturbative formulation of
string/M-theory in general backgrounds is
still lacking.
But there are recent successes for special backgrounds:
D-branes \cite{Polchinski:1995mt};
M(atrix) theory \cite{Banks:1996vh,Ishibashi:1996xs};
the AdS/CFT correspondence
\cite{Maldacena:1997re,Gubser:1998bc,Witten:1998qj};
and, the PP-wave limit
\cite{Metsaev:2001bj,Metsaev:2002re,Blau:2001ne,Berenstein:2002jq}.
Note that I include nontrivial semiclassical descriptions
under what I call ``nonperturbative.'' 
It is of considerable interest to study these
nonpertubative formulations in
relation to SYM on the lattice.
I next survey an interesting example
of the AdS/CFT correspondence.  In this example,
both the gauge and gravitational
theories are known.  Detailed information
on the IR physics of the gauge theory,
extracted from lattice studies, would
nicely supplement the many continuum results
that have been obtained.

\bfe{AdS/CFT correspondence.}
Here, the nontrivial SYM vacuum dynamics takes on a gravitational
meaning.  Everybody has heard about $AdS_5 \times S^5$.
A very interesting generalization is the 
Klebanov-Strassler (KS) construction \cite{Klebanov:2000hb}.
It is based on the Klebanov-Witten (KW) construction \cite{Klebanov:1998hh},
with a gravitational theory on $AdS_5 \times T^{1,1}$.
The KW dual gauge theory is an $\Ncal=1$ superconformal
field theory (SCFT), whose field content
and action is known.

\begin{figure}
\begin{center}
\includegraphics[width=3in,height=1.5in]{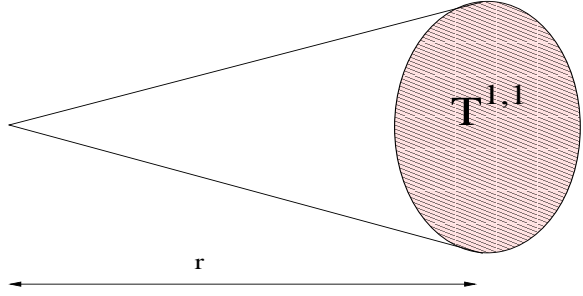}
\end{center}
\caption{A schematic view of the conifold.
\label{coni}}
\end{figure}

\begin{figure}
\begin{center}
\includegraphics[width=3in,height=1.2in]{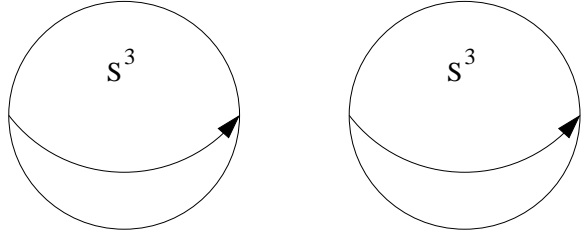}
\end{center}
\caption{The quotient of $S^3 \times S^3$ that
leads to $T^{1,1}$.
\label{t11q}}
\end{figure}

\bfe{The conifold.}
Type IIB string theory, and in particular its
supergravity limit, is formulated on the 10d space:
\beq
\text{4d Minkowski} \times \text{conifold}.
\eeq
The conifold is a 6d space possessing a cone-like geometry, with $T^{1,1}$ base,
as illustrated in Fig.~\ref{coni}.
The 5d compact space
$T^{1,1}$ is a quotient manifold:
\beq
T^{1,1} = [SU(2) \times SU(2)] / U(1).
\eeq
The ``$1,1$'' denotes the $U(1)$ quotient generated by:
\beq
H = (\s_3 \otimes \obf) +  (\obf \otimes \s_3).
\eeq
Fig.~\ref{t11q} illustrates the rotation, generated
by $H$, that identifies the points in $SU(2) \times SU(2)
\simeq S^3 \times S^3$ to form $T^{1,1}$.

\bfe{Warping spacetime.}
A stack of $N_c$ D3 branes is placed at the
tip of the confold, where the size of the base shrinks to zero.
I.e., $r^2 ds_{T^{1,1}}^2 \to 0$ as $r \to 0$.
The D3 branes are gravitating, so they backreact
on the geometry, warping it.  As a consequence,
not too far from the branes (the near-horizon limit), the geometry is
$AdS_5 \times T^{1,1}$.

\bfe{Killing spinors.}
One advantage of the conifold construction
is that Type IIB SUGRA on $AdS_5 \times T^{1,1}$
only preserves eight Killing spinors, whereas
Type IIB on $AdS_5 \times S^5$ preserves 32 Killing
spinors.\footnote{The number of Killing spinors
determines the amount of \susy\ in the
supergravity theory.}
In the dual gauge theory, we get an $\Ncal=1$ superconformal
field theory rather than $\Ncal=4$ superconformal
field theory.

\begin{figure}
\begin{center}
\includegraphics[width=3.5in,height=1.5in,bb=0 0 515 178,clip]{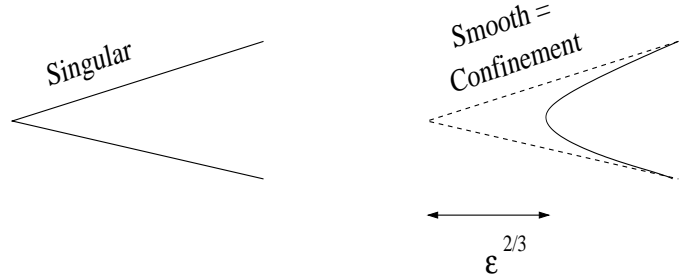}
\end{center}
\caption{The singularity resolution in
the deformed conifold.  The parameter $\e$
is used to describe the geometric deformation,
and makes an appearance in the glueball mass.
\label{KSres}}
\end{figure}

\bfe{The singularity at the tip and its
resolution.}
An $\Ncal=1$ SCFT is a much more promising start,
being closer to the real world.\footnote{$\Ncal=1$ is
necessary in order to have chiral representations.}
There is a singularity at the tip of the conifold, $r=0$.
In the dual gauge theory, this is reflected
by the absence of an IR cutoff.
Klebanov-Strassler resolve the singularity
with the deformed conifold, depicted
schematically in Fig.~\ref{KSres}.
They show that this is equivalent to confinement
in the IR of the dual gauge theory, due to
running of gauge couplings once conformal
symmetry (dual to the scaling isometry
of AdS) is broken.

On the gravity side of the duality, the deformation
of the conifold breaks half the Killing spinors.
On the gauge theory side, this breaks the
fermionic conformal charges, reducing to an
$\Ncal=1$ \susyc\ gauge theory that is not conformally
invariant.  Now we are ``very close'' to the real world;
i.e., warped, supersymmetric extensions to the
Standard Model.\footnote{For
a review of phenomenological models of this
type, see \cite{Gherghetta:2006ha}.}

\bfe{D7 probes and the equivalence to adding
quark flavors.}
Introduction of $N_f$ D7 probe branes allows
for a weakly coupled $U(N_f)$ gauge theory
in the dual gauge theory, due to volume suppression
associated with reduction to 4d:
\beq
g_f^2 \sim (\text{volume})^{-1}.
\eeq
The embedding for the D7 branes can generate bare
masses for ``quarks'' of dual gauge theory.
The low energy partons are really bound
states, very much like in technicolor.\footnote{
Unlike technicolor models, masses for the quarks
associated with probe D7 branes are allowed, 
because the spectrum is vector-like.}
Here, the $U(N_c)$ gauge theory
associated with the D3 branes 
plays the role of the technicolor-like group.

\begin{figure}
\begin{center}
\includegraphics[height=2.5in,width=2.5in]{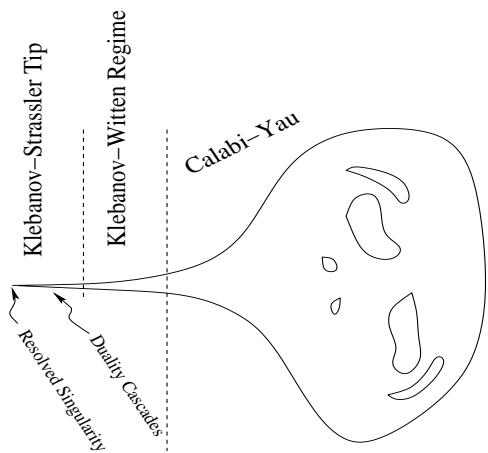}
\end{center}
\caption{A combination of a Klebanov-Strassler throat
and a Calabi-Yau ``boundary''.  This can be
thought of as a refinement of Randall-Sundrum 1 type models.
\label{rs1f}}
\end{figure}

\bfe{Noncompact $\to$ compact Calabi-Yau.}
Recently, studies of the low energy effective
4d and 5d theories derived from the probe
D7 branes have been conducted
\cite{Sakai:2003wu,Kuperstein:2004hy,Levi:2005hh,Gherghetta:2006yq}.
In the latter work \cite{Gherghetta:2006yq}, we also imagined a regulator
in the UV, following \cite{Giddings:2001yu}.
This occurs by capping off the conifold with
a compact Calabi-Yau manifold far from the tip,
as in Fig.~\ref{rs1f}.

\begin{figure}
\begin{center}
\includegraphics[width=3in,height=1.5in]{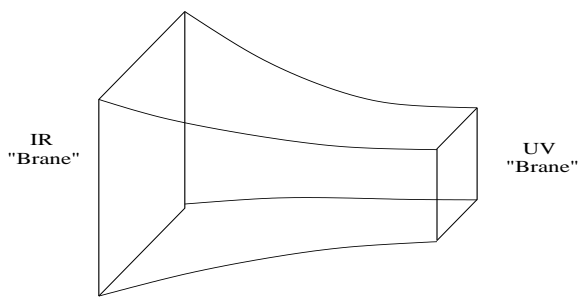}
\end{center}
\caption{A schematic picture of the
background geometry in the Randall-Sundrum 1 model \cite{Randall:1999ee}.
The figure is meant to indicate that,
at the IR brane, distances are dilated by the warp
factor of AdS.
\label{rsf2}}
\end{figure}

\bfe{Randall-Sundrum 1 interpretation.}
The Calabi-Yau and the KS tip are, respectively, to be
thought of as refinements of the UV and IR branes
of, say, Randall-Sundrum 1 type models, shown
schmatically in Fig.~\ref{rsf2}.  
The refined picture, which has already
been mentioned above, is shown in Fig.~\ref{rs1f}.
There is a potential---though challenging---interplay
between warped extra dimension models,
AdS/CFT, and lattice SYM.  A detailed study of the IR physics
of the dual gauge theory would have implications
for the gravitational picture.  Thus,
the goal is to be able to simulate
the gauge theory on the lattice.

\mys{$AdS_3/CFT_2$ and deconstructed (4,4) super-QCD}
There are many more generalizations of AdS/CFT.
In particular, $AdS_3/CFT_2$ may be accessible
through lattice studies.
The duality in the 2d (4,4)-supersymmetric QCD (the D1-D5 brane
intersection theory) 
is currently under study using a deconstruction lattice
\susy\ approach \cite{Giedt:2006dd}.
The (4,4) lattice with matter is a generalization
of the recent
(2,2) super-QCD constructions of Kaplan
and Endres \cite{Endres:2006ic}.
The (4,4) lattice theory has two exact supercharges
and only site, ordinary link, and diagonal link fields.

\mys{Conclusions}
Although it is challenging to write down
supersymmetric lattice field theories that
have a ``good'' quantum continuum limit, some
examples do exist.
A wealth of exciting applications await.
The most interesting are those related
to \susy\ breaking, quantum gravity
and the AdS/CFT correspondence.
Certain nonpertubative features, such
as spectra, can be studied on the lattice with results that
are of broad and current interest.
For practical reasons, and due to the existence of
very promising formulations, the 2d lattice
models could be the most fruitful numerically,
in the short term.

\section*{Acknowledgements}
Thanks are due to Erich Poppitz,
David B. Kaplan, Herbert Neuberger,
Pavlos Vranas, Alessandro D'Adda,
Noboru Kawamota, Simon Catterall, 
Falk Bruckmann 
and Mark de Kok for helpful communications.
This work was supported in part by the U.S.~Department of Energy
under grant No.~DE-FG02-94ER-40823.


\begin{thebibliography}{99}

\bibitem{Giedt:2006pd}
J.~Giedt,
``Deconstruction and other approaches to supersymmetric lattice field
theories,''
Int.\ J.\ Mod.\ Phys.\ A {\bf 21} (2006) 3039
[arXiv:hep-lat/0602007].


\bibitem{Catterall:2005eh}
S.~Catterall,
``Dirac-K\"ahler fermions and exact lattice supersymmetry,''
PoS {\bf LAT2005} (2006) 006
[arXiv:hep-lat/0509136].


\bibitem{Feo:2004kx}
A.~Feo,
``Predictions and recent results in SUSY on the lattice,''
Mod.\ Phys.\ Lett.\ A {\bf 19} (2004) 2387
[arXiv:hep-lat/0410012].

\bibitem{Feo:2002yi}
A.~Feo,
``Supersymmetry on the lattice,''
Nucl.\ Phys.\ Proc.\ Suppl.\  {\bf 119} (2003) 198
[arXiv:hep-lat/0210015].

\bibitem{Montvay:2001aj}
I.~Montvay,
``Supersymmetric Yang-Mills theory on the lattice,''
Int.\ J.\ Mod.\ Phys.\ A {\bf 17} (2002) 2377
[arXiv:hep-lat/0112007].

\bibitem{Catterall:2000rv}
S.~Catterall and E.~Gregory,
``A lattice path integral for supersymmetric quantum mechanics,''
Phys.\ Lett.\ B {\bf 487} (2000) 349
[arXiv:hep-lat/0006013].

\bibitem{Giedt:2004qs}
J.~Giedt and E.~Poppitz,
``Lattice supersymmetry, superfields and renormalization,''
JHEP {\bf 0409}, 029 (2004)
[arXiv:hep-th/0407135].

\bibitem{Wess:1973kz}
J.~Wess and B.~Zumino,
``A Lagrangian Model Invariant Under Supergauge Transformations,''
Phys.\ Lett.\ B {\bf 49} (1974) 52.

\bibitem{Elitzur:1983nj}
S.~Elitzur and A.~Schwimmer,
``N=2 Two-Dimensional Wess-Zumino Model On The Lattice,''
Nucl.\ Phys.\ B {\bf 226}, 109 (1983).

\bibitem{Sakai:1983dg}
N.~Sakai and M.~Sakamoto,
``Lattice Supersymmetry And The Nicolai Mapping,''
Nucl.\ Phys.\ B {\bf 229}, 173 (1983).

\bibitem{Reisz:1987da}
T.~Reisz,
``A Power Counting Theorem For Feynman Integrals On The Lattice,''
Commun.\ Math.\ Phys.\  {\bf 116} (1988) 81.

\bibitem{Beccaria:1998vi}
M.~Beccaria, G.~Curci and E.~D'Ambrosio,
``Simulation of supersymmetric models with a local Nicolai map,''
Phys.\ Rev.\ D {\bf 58} (1998) 065009
[arXiv:hep-lat/9804010].

\bibitem{Catterall:2001fr}
S.~Catterall and S.~Karamov,
``Exact lattice supersymmetry: the two-dimensional N = 2 Wess-Zumino  model,''
Phys.\ Rev.\ D {\bf 65} (2002) 094501
[arXiv:hep-lat/0108024].

\bibitem{Catterall:2001wx}
S.~Catterall and S.~Karamov,
``A two-dimensional lattice model with exact supersymmetry,''
Nucl.\ Phys.\ Proc.\ Suppl.\  {\bf 106} (2002) 935
[arXiv:hep-lat/0110071].

\bibitem{Giedt:2005ae}
J.~Giedt,
``R-symmetry in the Q-exact (2,2) 2d lattice Wess-Zumino model,''
Nucl.\ Phys.\ B {\bf 726} (2005) 210
[arXiv:hep-lat/0507016].

\bibitem{Curci:1986sm}
G.~Curci and G.~Veneziano,
``Supersymmetry And The Lattice: A Reconciliation?,''
Nucl.\ Phys.\ B {\bf 292} (1987) 555.

\bibitem{Narayanan:1992wx}
R.~Narayanan and H.~Neuberger,
``Infinitely many regulator fields for chiral fermions,''
Phys.\ Lett.\ B {\bf 302} (1993) 62
[arXiv:hep-lat/9212019].

\bibitem{Narayanan:1993sk}
R.~Narayanan and H.~Neuberger,
``Chiral determinant as an overlap of two vacua,''
Nucl.\ Phys.\ B {\bf 412} (1994) 574
[arXiv:hep-lat/9307006].

\bibitem{Neuberger:1997fp}
H.~Neuberger,
``Exactly massless quarks on the lattice,''
Phys.\ Lett.\ B {\bf 417} (1998) 141
[arXiv:hep-lat/9707022].

\bibitem{Maru:1997kh}
N.~Maru and J.~Nishimura,
``Lattice formulation of supersymmetric 
Yang-Mills theories without fine-tuning,''
Int.\ J.\ Mod.\ Phys.\ A {\bf 13} (1998) 2841
[arXiv:hep-th/9705152].

\bibitem{Neuberger:1997bg}
H.~Neuberger,
``Vector like gauge theories with almost massless fermions on the  lattice,''
Phys.\ Rev.\ D {\bf 57} (1998) 5417
[arXiv:hep-lat/9710089].

\bibitem{Kaplan:1992bt}
D.~B.~Kaplan,
``A Method for simulating chiral fermions on the lattice,''
Phys.\ Lett.\ B {\bf 288} (1992) 342
[arXiv:hep-lat/9206013].

\bibitem{Nishimura:1997vg}
J.~Nishimura,
``Four-dimensional N = 1 supersymmetric Yang-Mills theory on the lattice
without fine-tuning,''
Phys.\ Lett.\ B {\bf 406} (1997) 215
[arXiv:hep-lat/9701013].

\bibitem{Kaplan:1999jn}
D.~B.~Kaplan and M.~Schmaltz,
``Supersymmetric Yang-Mills theories from domain wall fermions,''
Chin.\ J.\ Phys.\  {\bf 38} (2000) 543
[arXiv:hep-lat/0002030].
  
\bibitem{Fleming:2000fa}
G.~T.~Fleming, J.~B.~Kogut and P.~M.~Vranas,
``Super Yang-Mills on the lattice with domain wall fermions,''
Phys.\ Rev.\ D {\bf 64} (2001) 034510
[arXiv:hep-lat/0008009].

\bibitem{Veneziano:1982ah}
G.~Veneziano and S.~Yankielowicz,
``An Effective Lagrangian For The Pure N=1 Supersymmetric Yang-Mills
Theory,''
Phys.\ Lett.\ B {\bf 113} (1982) 231.

\bibitem{Farrar:1997fn}
G.~R.~Farrar, G.~Gabadadze and M.~Schwetz,
``On the effective action of N = 1 supersymmetric Yang-Mills theory,''
Phys.\ Rev.\ D {\bf 58} (1998) 015009
[arXiv:hep-th/9711166].

\bibitem{Farrar:1998rm}
G.~R.~Farrar, G.~Gabadadze and M.~Schwetz,
``The spectrum of softly broken N = 1 supersymmetric Yang-Mills theory,''
Phys.\ Rev.\ D {\bf 60} (1999) 035002
[arXiv:hep-th/9806204].

\bibitem{Gabadadze:1998bi}
G.~Gabadadze,
``The discrete Z(2N(c)) symmetry and effective superpotential in SUSY
gluodynamics,''
Nucl.\ Phys.\ B {\bf 544} (1999) 650
[arXiv:hep-th/9808005].

\bibitem{Cerdeno:2003us}
D.~G.~Cerdeno, A.~Knauf and J.~Louis,
``A note on effective N = 1 super Yang-Mills theories versus lattice
results,''
Eur.\ Phys.\ J.\ C {\bf 31} (2003) 415
[arXiv:hep-th/0307198].

\bibitem{Merlatti:2004df}
P.~Merlatti and F.~Sannino,
``Extending the Veneziano-Yankielowicz effective theory,''
Phys.\ Rev.\ D {\bf 70} (2004) 065022
[arXiv:hep-th/0404251].
  
\bibitem{Auzzi:2005fi}
R.~Auzzi and F.~Sannino,
``Hidden modulus in the extended Veneziano-Yankielowicz theory,''
Phys.\ Rev.\ D {\bf 72} (2005) 105003
[arXiv:hep-th/0509166].

\bibitem{Vranas:2006zk}
P.~M.~Vranas,
``Gap domain wall fermions,''
Phys.\ Rev.\ D {\bf 74} (2006) 034512
[arXiv:hep-lat/0606014]; 
this proceedings,
``The method of gap domain-wall fermions,''
PoS(LAT2006)059.
\bibitem{Arkani-Hamed:2001ca}
N.~Arkani-Hamed, A.~G.~Cohen and H.~Georgi, 
``(De)constructing
dimensions,'' 
Phys.\ Rev.\ Lett.\  {\bf 86}, 4757 (2001)
[arXiv:hep-th/0104005];

\bibitem{Hill:2000mu}
C.~T.~Hill, S.~Pokorski and J.~Wang, 
``Gauge invariant effective
Lagrangian for Kaluza-Klein modes,'' 
Phys.\ Rev.\ D {\bf 64},
105005 (2001) [arXiv:hep-th/0104035].

\bibitem{Kaplan:2002wv}
D.~B.~Kaplan, E.~Katz and M.~Unsal,
``Supersymmetry on a spatial lattice,''
JHEP {\bf 0305} (2003) 037
[arXiv:hep-lat/0206019].

\bibitem{Cohen:2003xe}
A.~G.~Cohen, D.~B.~Kaplan, E.~Katz and M.~Unsal,
``Supersymmetry on a Euclidean spacetime lattice. I:
A target theory with  four supercharges,''
JHEP {\bf 0308}, 024 (2003)
[arXiv:hep-lat/0302017].

\bibitem{Cohen:2003qw}
A.~G.~Cohen, D.~B.~Kaplan, E.~Katz and M.~Unsal,
``Supersymmetry on a Euclidean spacetime lattice. II: 
Target theories with eight supercharges,''
JHEP {\bf 0312}, 031 (2003)
[arXiv:hep-lat/0307012].

\bibitem{Kaplan:2005ta}
D.~B.~Kaplan and M.~Unsal,
``A Euclidean lattice construction of supersymmetric Yang-Mills theories
with sixteen supercharges,''
JHEP {\bf 0509} (2005) 042
[arXiv:hep-lat/0503039].

\bibitem{Endres:2006ic}
M.~G.~Endres and D.~B.~Kaplan,
``Lattice formulation of (2,2) supersymmetric gauge theories with matter
fields,''
arXiv:hep-lat/0604012.

\bibitem{Giedt:2006dd}
J.~Giedt,
``A deconstruction lattice description of 
the D1/D5 brane world-volume gauge theory,''
arXiv:hep-lat/0605004.

\bibitem{Onogi:2005cz}
T.~Onogi and T.~Takimi,
``Perturbative study of the supersymmetric lattice theory from matrix model,''
Phys.\ Rev.\ D {\bf 72} (2005) 074504
[arXiv:hep-lat/0506014].

\bibitem{Giedt:2003ve}
J.~Giedt,
``Non-positive fermion determinants in lattice supersymmetry,''
Nucl.\ Phys.\ B {\bf 668} (2003) 138
[arXiv:hep-lat/0304006].

\bibitem{Giedt:2003vy}
J.~Giedt,
``The fermion determinant in (4,4) 2d lattice super-Yang-Mills,''
Nucl.\ Phys.\ B {\bf 674} (2003) 259
[arXiv:hep-lat/0307024].

\bibitem{Giedt:2004tn}
J.~Giedt,
``Deconstruction, 2d lattice super-Yang-Mills, and the dynamical lattice
spacing,''
arXiv:hep-lat/0405021.

\bibitem{D'Adda:2004jb}
A.~D'Adda, I.~Kanamori, N.~Kawamoto and K.~Nagata,
``Twisted superspace on a lattice,''
Nucl.\ Phys.\ B {\bf 707} (2005) 100
[arXiv:hep-lat/0406029].

\bibitem{D'Adda:2005zk}
A.~D'Adda, I.~Kanamori, N.~Kawamoto and K.~Nagata,
``Exact extended supersymmetry on a lattice: Twisted N = 2 super  Yang-Mills
in two dimensions,''
Phys.\ Lett.\ B {\bf 633} (2006) 645
[arXiv:hep-lat/0507029].

\bibitem{Bruckmann:2006ub}
F.~Bruckmann and M.~de Kok,
``Noncommutativity approach to supersymmetry on the lattice: SUSY quantum
mechanics and an inconsistency,''
Phys.\ Rev.\ D {\bf 73} (2006) 074511
[arXiv:hep-lat/0603003].

\bibitem{Bruckmann:2006kb}
F.~Bruckmann, S.~Catterall and M.~de Kok,
``A critique of the link approach to exact lattice supersymmetry,''
arXiv:hep-lat/0611001.

\bibitem{Sugino:2003yb}
F.~Sugino,
``A lattice formulation of super Yang-Mills theories with exact
supersymmetry,''
JHEP {\bf 0401} (2004) 015
[arXiv:hep-lat/0311021].

\bibitem{Sugino:2004qd}
F.~Sugino,
``Super Yang-Mills theories on the two-dimensional lattice with exact
supersymmetry,''
JHEP {\bf 0403} (2004) 067
[arXiv:hep-lat/0401017].

\bibitem{Sugino:2004uv}
F.~Sugino,
``Various super Yang-Mills theories with exact supersymmetry on the  lattice,''
JHEP {\bf 0501} (2005) 016
[arXiv:hep-lat/0410035].

\bibitem{Sugino:2006uf}
F.~Sugino,
``Two-dimensional compact N = (2,2) lattice super Yang-Mills theory with
exact supersymmetry,''
Phys.\ Lett.\ B {\bf 635} (2006) 218
[arXiv:hep-lat/0601024].

\bibitem{Catterall:2004np}
S.~Catterall,
``A geometrical approach to N = 2 super Yang-Mills theory on the two
dimensional lattice,''
JHEP {\bf 0411} (2004) 006
[arXiv:hep-lat/0410052].

\bibitem{Catterall:2006jw}
S.~Catterall,
``Simulations of $\Ncal = 2$ super Yang-Mills theory in two dimensions,''
JHEP {\bf 0603} (2006) 032
[arXiv:hep-lat/0602004];
``On the restoration of supersymmetry in twisted two-dimensional lattice
Yang-Mills theory,''
arXiv:hep-lat/0612008.

\bibitem{Bonini:2004pm}
M.~Bonini and A.~Feo,
``Wess-Zumino model with exact supersymmetry on the lattice,''
JHEP {\bf 0409} (2004) 011
[arXiv:hep-lat/0402034].

\bibitem{Bonini:2004wv}
M.~Bonini and A.~Feo,
``Exact supersymmetry on the lattice: The Wess-Zumino model,''
Nucl.\ Phys.\ Proc.\ Suppl.\  {\bf 140}, 766 (2005)
[arXiv:hep-lat/0409068].
  
\bibitem{Bonini:2005qx}
M.~Bonini and A.~Feo,
``Exact lattice Ward-Takahashi identity for the N = 1 Wess-Zumino model,''
Phys.\ Rev.\ D {\bf 71} (2005) 114512
[arXiv:hep-lat/0504010].

\bibitem{Catterall:2003uf}
S.~Catterall and S.~Ghadab,
``Lattice sigma models with exact supersymmetry,''
JHEP {\bf 0405} (2004) 044
[arXiv:hep-lat/0311042].

\bibitem{Ghadab:2004rt}
S.~Ghadab,
``Lattice study of the O(3) supersymmetric sigma model,''
Nucl.\ Phys.\ Proc.\ Suppl.\  {\bf 140} (2005) 760
[arXiv:hep-lat/0409102].

\bibitem{Catterall:2006sj}
S.~Catterall and S.~Ghadab,
``Twisted supersymmetric sigma model on the lattice,''
arXiv:hep-lat/0607010.

\bibitem{Affleck:1982as}
I.~Affleck, J.~A.~Harvey and E.~Witten,
``Instantons And (Super)Symmetry Breaking In (2+1)-Dimensions,''
Nucl.\ Phys.\ B {\bf 206} (1982) 413.

\bibitem{Polchinski:1995mt}
J.~Polchinski,
``Dirichlet-Branes and Ramond-Ramond Charges,''
Phys.\ Rev.\ Lett.\  {\bf 75} (1995) 4724
[arXiv:hep-th/9510017].

\bibitem{Banks:1996vh}
T.~Banks, W.~Fischler, S.~H.~Shenker and L.~Susskind,
``M theory as a matrix model: A conjecture,''
Phys.\ Rev.\ D {\bf 55} (1997) 5112
[arXiv:hep-th/9610043].

\bibitem{Ishibashi:1996xs}
N.~Ishibashi, H.~Kawai, Y.~Kitazawa and A.~Tsuchiya,
``A large-N reduced model as superstring,''
Nucl.\ Phys.\ B {\bf 498} (1997) 467
[arXiv:hep-th/9612115].

\bibitem{Maldacena:1997re}
J.~M.~Maldacena,
``The large N limit of superconformal field theories and supergravity,''
Adv.\ Theor.\ Math.\ Phys.\  {\bf 2} (1998) 231
[Int.\ J.\ Theor.\ Phys.\  {\bf 38} (1999) 1113]
[arXiv:hep-th/9711200].

\bibitem{Gubser:1998bc}
S.~S.~Gubser, I.~R.~Klebanov and A.~M.~Polyakov,
``Gauge theory correlators from non-critical string theory,''
Phys.\ Lett.\ B {\bf 428} (1998) 105
[arXiv:hep-th/9802109].

\bibitem{Witten:1998qj}
E.~Witten,
``Anti-de Sitter space and holography,''
Adv.\ Theor.\ Math.\ Phys.\  {\bf 2} (1998) 253
[arXiv:hep-th/9802150].

\bibitem{Metsaev:2001bj}
R.~R.~Metsaev,
``Type IIB Green-Schwarz superstring in plane wave Ramond-Ramond  background,''
Nucl.\ Phys.\ B {\bf 625} (2002) 70
[arXiv:hep-th/0112044].

\bibitem{Metsaev:2002re}
R.~R.~Metsaev and A.~A.~Tseytlin,
``Exactly solvable model of superstring in plane wave Ramond-Ramond
background,''
Phys.\ Rev.\ D {\bf 65} (2002) 126004
[arXiv:hep-th/0202109].

\bibitem{Blau:2001ne}
M.~Blau, J.~Figueroa-O'Farrill, C.~Hull and G.~Papadopoulos,
``A new maximally supersymmetric background of IIB superstring theory,''
JHEP {\bf 0201} (2002) 047
[arXiv:hep-th/0110242].

\bibitem{Berenstein:2002jq}
D.~Berenstein, J.~M.~Maldacena and H.~Nastase,
``Strings in flat space and pp waves from N = 4 super Yang Mills,''
JHEP {\bf 0204} (2002) 013
[arXiv:hep-th/0202021].

\bibitem{Klebanov:2000hb}
I.~R.~Klebanov and M.~J.~Strassler,
``Supergravity and a confining gauge theory: Duality cascades and
$\chi$SB-resolution of naked singularities,''
JHEP {\bf 0008} (2000) 052
[arXiv:hep-th/0007191].

\bibitem{Klebanov:1998hh}
I.~R.~Klebanov and E.~Witten,
``Superconformal field theory on threebranes at a Calabi-Yau  singularity,''
Nucl.\ Phys.\ B {\bf 536} (1998) 199
[arXiv:hep-th/9807080].

\bibitem{Gherghetta:2006ha}
For a review, see T.~Gherghetta,
``Les Houches lectures on warped models and holography,''
arXiv:hep-ph/0601213.

\bibitem{Sakai:2003wu}
T.~Sakai and J.~Sonnenschein,
``Probing flavored mesons of confining gauge theories by supergravity,''
JHEP {\bf 0309} (2003) 047
[arXiv:hep-th/0305049].

\bibitem{Kuperstein:2004hy}
S.~Kuperstein,
``Meson spectroscopy from holomorphic probes on the warped deformed
conifold,''
JHEP {\bf 0503} (2005) 014
[arXiv:hep-th/0411097].

\bibitem{Levi:2005hh}
T.~S.~Levi and P.~Ouyang,
``Mesons and flavor on the conifold,''
arXiv:hep-th/0506021.

\bibitem{Gherghetta:2006yq}
T.~Gherghetta and J.~Giedt,
``Bulk fields in AdS(5) from probe D7 branes,''
Phys.~Rev. D {\bf 74} (2006) 066007 [hep-th/0605212].

\bibitem{Giddings:2001yu}
S.~B.~Giddings, S.~Kachru and J.~Polchinski,
``Hierarchies from fluxes in string compactifications,''
Phys.\ Rev.\ D {\bf 66} (2002) 106006
[arXiv:hep-th/0105097].

\bibitem{Randall:1999ee}
L.~Randall and R.~Sundrum,
``A large mass hierarchy from a small extra dimension,''
Phys.\ Rev.\ Lett.\  {\bf 83} (1999) 3370
[arXiv:hep-ph/9905221].

\end{thebibliography}
\end{document}